\documentstyle[11pt,IAU207_pasp,twoside,psfig]{article}
\def\edcomment#1{\iffalse\marginpar{\raggedright\sl#1\/}\else\relax\fi}
\reversemarginpar

\begin{document}
\title{Second Parameter Effects in and between M3 and Palomar~3}
\author{M. Catelan, R. T. Rood}
\affil{University of Virginia,
       Department of Astronomy,
       P.O.~Box~3818,
       Charlottesville, VA 22903-0818, USA}
\author{F. R. Ferraro}
\affil{Osservatorio Astronomico di Bologna,
       via Ranzani~1,
       I-40126 Bologna, Italy}

\begin{abstract}
We study the globular clusters M3 and Palomar~3 as a 
``second parameter (2ndP) pair," showing that:  
i)~M3 has a surprisingly strong {\em internal} 2ndP; 
ii)~The dispersion in mass on the Pal~3 horizontal branch (HB) 
is intrinsically very small, leading to the most 
apparent differences in HB morphology between M3 and Pal~3;  
iii)~Ignoring the difference in HB mass dispersion between M3 and 
Pal~3, their relative HB types can be accounted for by a fairly 
small difference in age, of order 0.5--1~Gyr. 
\end{abstract}

\begin{figure}[ht]	
\plotfiddle{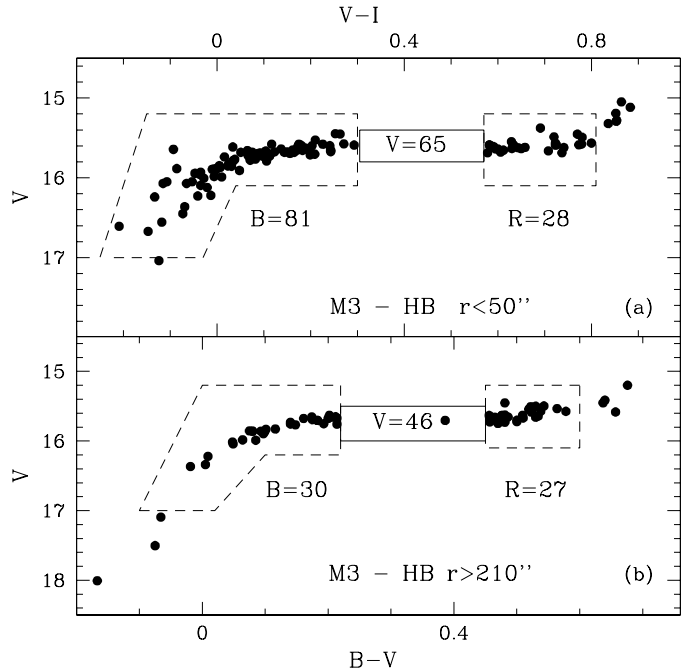}{9cm}{0}{50}{50}{-150}{-85}
\caption{M3, once thought to be the prototypical ``canonical" globular 
    cluster, shows a very strong internal second parameter, with a much bluer 
    HB morphology in its inner regions (upper diagram) than in its outskirts 
    (lower diagram).}
\end{figure}

\section{Internal Second Parameter in M3}

Figure~1 shows the M3 HB CMD for the 
innermost ($r < 50\arcsec$, upper panel) and for the 
outermost  ($r > 210\arcsec$; lower panel) cluster regions. 
The numbers of red (``R"), RR Lyrae variable (``V") and 
blue (``B") HB stars 
are indicated. It is apparent that {\em the innermost 
regions of M3 have a much bluer HB than the outermost 
ones}. Even though the inner regions were 
observed with HST-WFPC2 while the outermost regions were observed 
from the ground, we emphasize that the difference is real, 
extremely significant, and {\em cannot} be accounted for by 
any conceivable source of systematic observational error.

\section{An Intrinsically Small Mass Dispersion on the Pal~3 ZAHB}

The HB region of the CMD of Pal~3 is shown in Fig.~2. This is 
essentially the same as the Stetson et al. (1999) HST-WFPC2 CMD, 
except that here the individual RR Lyrae variables are plotted, 
based on mean colors and magnitudes derived by combining the HST 
photometry and the Borissova et al. (1998) ground-based photometry. 
Note that {\em all} cluster RR Lyraes are ab-type variables, and 
there are no blue-HB stars. A theoretical ZAHB and evolutionary 
track are overplotted. The single track spans the entire observed
HB showing that little (if any) intrinsic dispersion in mass on the 
ZAHB is needed to account for the Pal~3 HB morphology. Extensive 
Monte Carlo simulations confirm this.

\section{The Difference in Age between M3 and Pal~3}
Ignoring, for the sake of argument, the internal 2ndP in
M3 and the difference in mass dispersion between the M3 and Pal~3 HBs, 
we find, using an approach similar to Catelan's (2000), that the relative 
HB types of M3 and Pal~3 can be easily accounted for by a difference in 
age of $\approx 0.5-1$~Gyr (Fig.~3), as indeed favored by VandenBerg 
(2000) from analysis of the clusters' turnoffs.
The larger age difference favored by Stetson et al. (1999), 
$\approx 2$~Gyr, is not consistent with the relative HB types of 
the clusters under the assumption that this is a ``bona fide"
2ndP pair.

\begin{figure}[ht]	
\plotfiddle{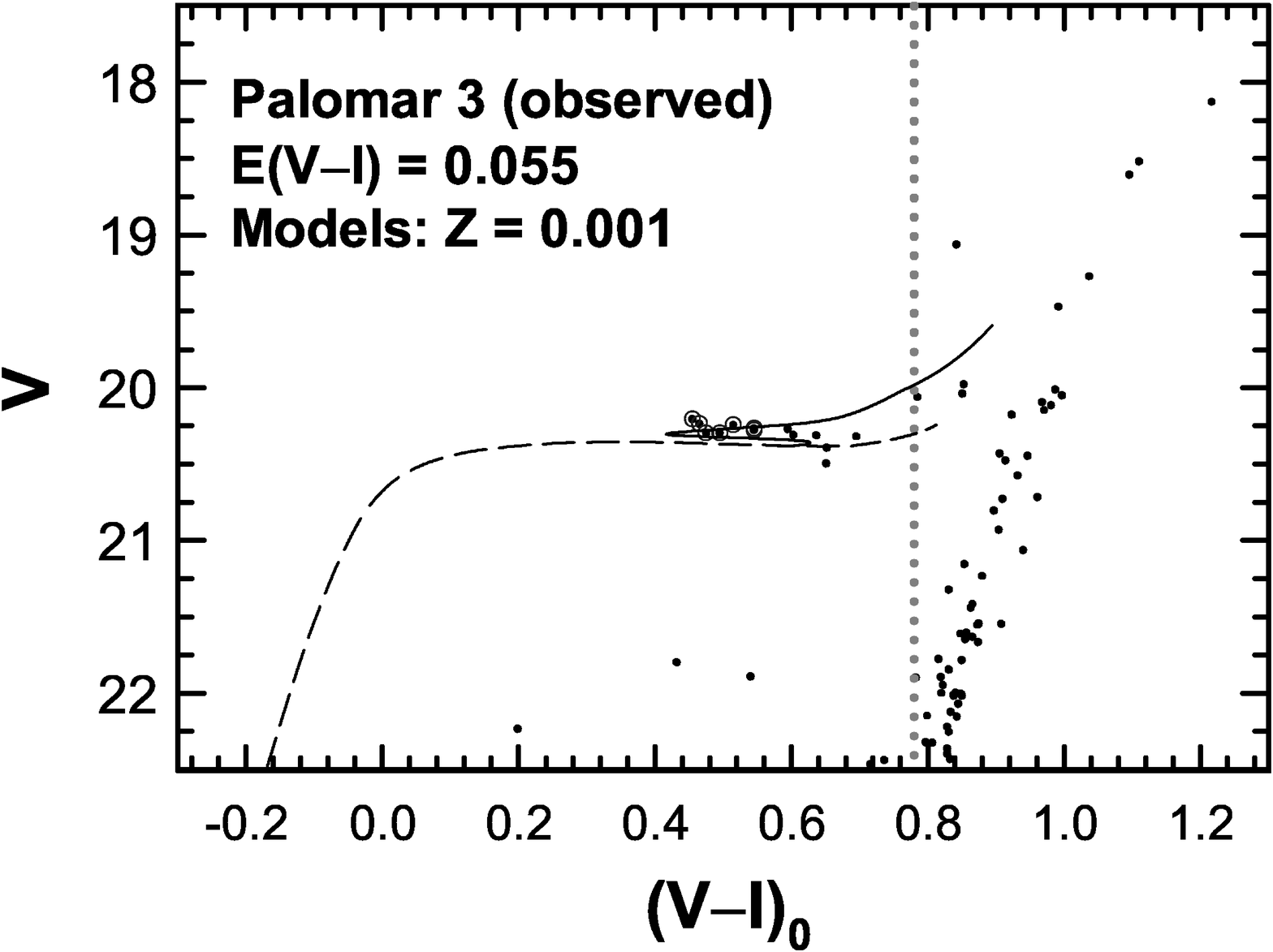}{8cm}{0}{13.5}{13.5}{-162}{0}
\caption{The CMD of Pal~3, with individual RR Lyrae variables plotted.
    A ZAHB and evolutionary track for $Z = 0.001$ are overplotted for 
    reference. The vertical dotted line shows the mean color of 
    the Pal~4/Eridanus HBs. Note that the HB morphology of Pal~3 can be 
    entirely accounted for without the need to assume a significant
    dispersion in mass on the ZAHB.}
\plotfiddle{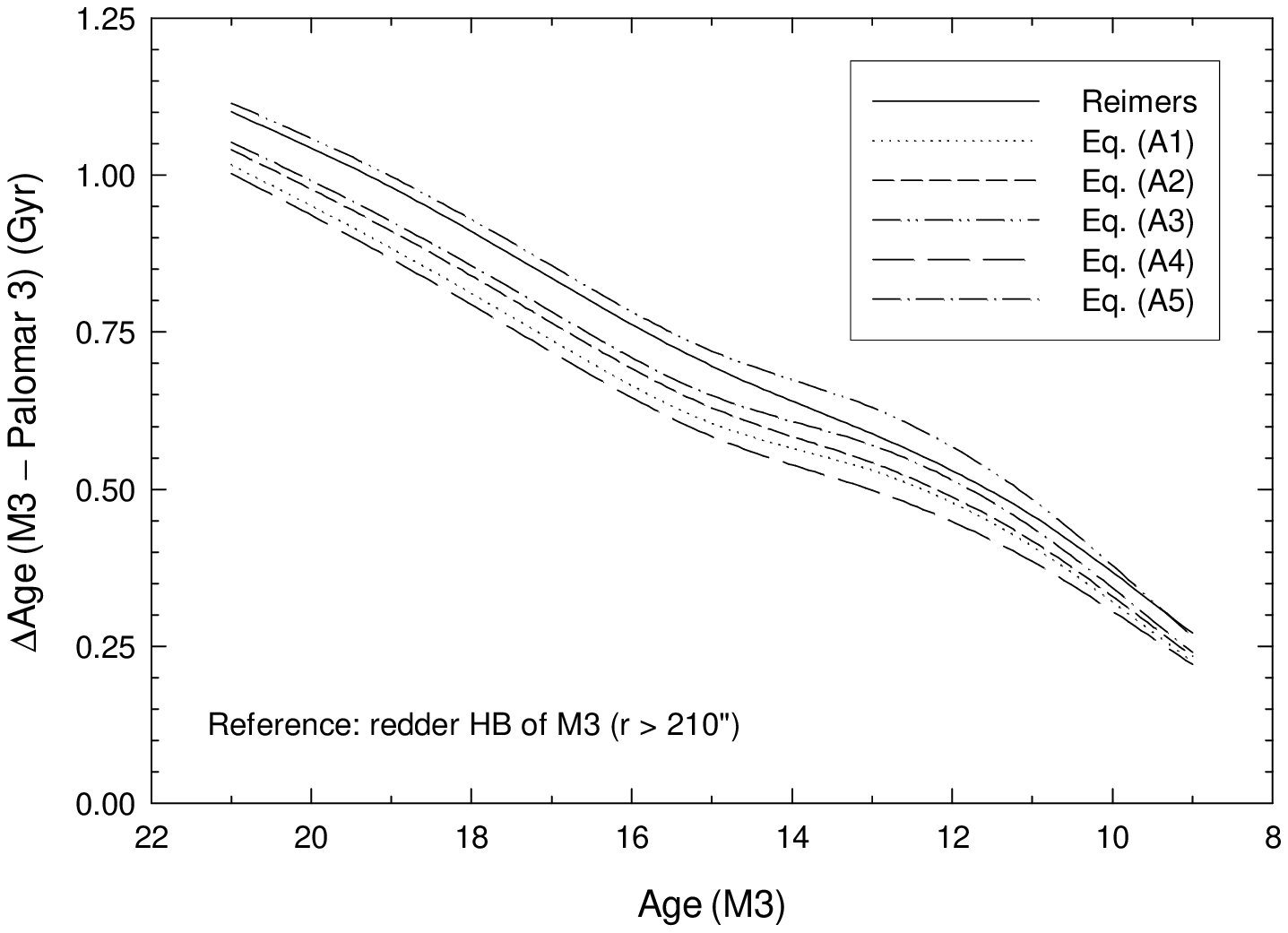}{9cm}{0}{75}{75}{-175}{0}
\caption{Relative HB morphology ages for M3 vs. Pal~3, plotted as a 
    function of the absolute M3 age. This assumes the outermost regions 
    of M3, with a redder HB, to be more representative of the cluster 
    as a whole. The different lines correspond to different mass 
    loss formulae for red giants (Catelan 2000).}
\end{figure}


\begin{references}
  \reference Borissova, J., Spassova, N., Catelan, M., \& Ivanov, V. D.
     1998, in ASP Conf. Ser. Vol.
     135, A Half-Century of Stellar Pulsation Interpretations,
     ed. P. A. Bradley \& J. A. Guzik (San Francisco: ASP), 188

  \reference Catelan, M. 2000, \apj, 531, 826

  \reference Stetson, P. B., et al. 1999, \aj, 117, 247

  \reference VandenBerg, D. A. 2000, \apjs, 129, 315
\end{references}
\end{document}